# Analytical and Numerical Approaches to Pricing the Path-Dependent Options with Stochastic Volatility

Kuperin Yu.A., Poloskov P.A.

Department of Physics, Saint Petersburg State University, Saint Petersburg, Russia

**Abstract**

In this paper new analytical and numerical approaches to valuating path-dependent options of European type have been developed. The model of stochastic volatility as a basic model has been chosen. For European options we could improve the path integral method, proposed B. Baaquie, and generalized it to the case of path-dependent options, where the payoff function depends on the history of changes in the underlying asset. The dependence of the implied volatility on the parameters of the stochastic volatility model has been studied. It is shown that with proper choice of model parameters one can accurately reproduce the actual behavior of implied volatility. As a consequence, it can assess more accurately the value of options. It should be noted that the methods developed here allow evaluating options with any payoff function.

## 1. Introduction

The classical theory of option pricing is based on the results obtained in 1973 in the Black, Scholes and Merton [1,2]. Their papers were based mainly on the assumption that the price of an asset subject to the dynamics of a specific stochastic process (geometric Brownian motion), and, consequently, has lognormal distribution. In the case of an efficient market with no arbitrage opportunities, and the constant volatility it turned out that the price of any option satisfies the differential equations in partial derivatives, which is known as the equation of Black-Scholes-Merton (BSM) For the case of simple vanilla options, the equation has an exact solution is known as the Black-Scholes formula [3,4]. The massive use of this formula showed that it does not always accurately describes the real option prices, in particular, it ignores the effect of «volatility smile» [3,4]. To solve the "volatility smile" problem several methods have been proposed. The main ones are the method of stochastic volatility [5,6].

In the case of stochastic volatility BSM equation becomes an equation Merton-Garman [5,6,7,8]. The increasing complexity of models describing the dynamics of options prices requires the use of new, more sophisticated mathematical methods, many of which are used in theoretical physics. In the case of stochastic volatility for European vanilla options, Heston using the Fourier transform received the exact formula [5]. An alternative way of solving the Merton-Garman equation can be obtained using the technique of path integral [9]. The idea of using path integral arises from the fact that the BSM equation can be written in a form similar to the Schrodinger equation for the wave function in quantum mechanics. For stock options were found exact expressions for the Hamiltonian and Lagrangian in this approach [10].

Path integral method is very useful for options with a complex payoff function, for example, when the payoff function depends on the values of the underlying asset for the duration of an option contract (path-dependent option). Examples of studies in this area are presented in the article [11]. These results contributed to the wide application of methods of theoretical physics in finance, and, in particular, the use of path integral for obtaining the exact or numerical solutions of the options pricing problems. In this regard, there is a need to develop flexible methods that could solve problems in the framework of stochastic volatility models and representations of the option price as a path integral.

Such techniques may be useful in practice, because they can more accurately assess the value of option contracts.

## 2. Stochasic models for underlying and volatility

First, we define the stochastic processes that underlie the model. For the dynamics of the underlying asset, we choose the Bachelier process:

$$dS = \mu S dt + \sigma S dZ_1 \quad (0.1)$$

As a stochastic process for volatility, one can choose various processes, for example,

$$dV = \mu V dt + \xi V dZ_2$$

$$dV = k(Q - V)dt + \xi V^{1/2} dZ_2$$

where $\mu > 0$ is the drift parameter, $\xi > 0$ is the so calld volatility of volatility, $Z_2$ is the Wiener process, $Q$ is the long volatility, $k$ is rate of return to the volatility mean.

These models were discussed in articles [7,8], respectively. Review and comparison of these models can be found in [6]. In this paper we use the general type of stochastic process for the dynamics of volatility:

$$dV = (\lambda + kV)dt + \xi V^\alpha dZ_2 \quad (0.2)$$

Here V= $\sigma^2$, $k$ is rate of return to the volatility mean, $\lambda$ is the long volatility, $\xi > 0$ is, the volatility of volatility $\alpha$ is some model parameter, $Z_2$ is the Wiener process, having the correlation coefficient $\rho$ with the Wiener process $Z_1$. It should be noted that the parameters $\lambda$ and $k$ should be chosen so that at any given time $t$, inequality $V > 0$ must be satisfied.

## 3. The option price as an path integral

The problem of finding the value of the option may present be using the standard methods in the form of differential equations in partial derivatives:

$$rf = \frac{df}{dt} + rS\frac{df}{dS} + (\lambda^* + \mu V)\frac{df}{dV} + \frac{1}{2}\xi^2 V^{2\alpha}\frac{d^2f}{dV^2} + \frac{1}{2}\sigma^2 S^2 \frac{d^2f}{dS^2} + \\ + \rho\sigma\xi SV^\alpha \frac{d^2f}{dVdS} \quad (0.3)$$

This equation is known as the equation of Merton-Garman. In this equation quite standard notations are used. Here $f$ is the price of the option, $r$ is risk-free interest rate, $\lambda^*$ is the renormalized long volatility.

One way of solving the Merton-Garman equation is a method in which this equation reduces to the path integral [9]. This method is described in detail in articles by B. Baquie [7,8].

In order to expand the domain of variables $0 \leq S < \infty, 0 \leq V < \infty$ in equation (1.3) to the whole real axis we make in (1.3) the following change of variables $S = e^x, V = e^y$, where $-\infty < x < \infty, -\infty < y < \infty,$.

It should be noted that below all further consideration will be in terms of reciprocal time $\tau = T - t$. After these changes of variables we obtain:

$$-\frac{df}{d\tau} + (r - \frac{1}{2}e^y)\frac{df}{dx} + (\lambda e^{-y} + \mu - \frac{1}{2}\xi^2 e^{2y(\alpha-1)})\frac{df}{dy} + \\ + \frac{1}{2}\xi^2 e^{2y(\alpha-1)}\frac{d^2f}{dy^2} + \frac{1}{2}e^y\frac{d^2f}{dx^2} + \rho\xi e^{y(\alpha-\frac{1}{2})}\frac{d^2f}{dxdy} = rf \quad (0.4)$$

This equation for arbitrary $\alpha$ can always be solved numerically. For the $\alpha = 1/2$ Heston [5] obtained an analytical expression using the Fourier transform. When $\alpha = 1$, the exact solution by using path integral was obtained by B. Baaquie [8].

Represent now the equation obtained in the form of the Schrodinger equation:

$$-\frac{\partial f}{\partial t} = (r + \widehat{H})f, \quad (0.5)$$

where $\widehat{H}$ is the analogue of the Hamiltonian operator in the Schrodinger equation and has the form:

$$\widehat{H}(x,y) = -(r - \frac{1}{2}e^y)\frac{d}{dx} - (\lambda e^{-y} + \mu - \frac{1}{2}\xi^2 e^{2y(\alpha-1)})\frac{d}{dy} - \frac{1}{2}\xi^2 e^{2y(\alpha-1)}\frac{d^2}{dy^2} - \frac{1}{2}e^y\frac{d^2}{dx^2} - \rho\xi e^{y(\alpha-\frac{1}{2})}\frac{d^2}{dxdy} \quad (0.6)$$

Note that in contrast to quantum mechanics, the above Hamiltonian operator is not selfadjoint. This, however, will not affect the validity of the subsequent calculations. The solution of equation (1.5) will have the following formal form

$$f(\tau) = e^{-\tau(r+\widehat{H})} f(0), \quad (0.7)$$

where $f(0)$ is a payoff function. For convenience, we rewrite equation (1.7) in the Dirac notation:

$$|f,\tau> = e^{-\tau(r+\widehat{H})} |f,0> \quad (0.8)$$

Suppose $|f,0> \equiv |g>$, as well as using the fact that $f(x,y) = <x,y|f>$ we write equation (1.8) in the form

$$f(x,y,\tau) = <x,y|e^{-\tau(r+\widehat{H})}|g> = e^{-r\tau}<x,y|e^{-\tau\widehat{H}}|g>, \quad (0.9)$$

Next one needs to use the completeness relation:

$$I = \int_{-\infty}^{\infty} dxdy |x,y><x,y|$$

and substitute this relation into equation (1.9). Then we obtain:

$$f(x,y,\tau) = e^{-r\tau}\int_{-\infty}^{\infty} dx'dy' <x,y|e^{-\tau\widehat{H}}|x',y'> g(x') \quad (0.10)$$

We have the integral equation for the value of an option of European type with a vanilla payoff function. In this equation the quantity $<x,y|e^{-\tau\widehat{H}}|x',y'>$ is called the propagator, and determines the transition probability from state $|x',y'>$ to state $|x,y>$. Now consider the integral on the right side of (1.10) in the form of a discrete path integral. For this, first divide the time $\tau$ into $n+1$ interval $\varepsilon = \tau/(n+1)$. Next, we substitute in equation (1.10) the completeness relation $n$ time and obtain

$$f(x,y,\tau) = e^{-r\tau}\int_{-\infty}^{\infty} dx_n dy_n ... dx_0 dy_0 <x_{n+1},y_{n+1}|e^{-\varepsilon\widehat{H}}|x_n,y_n> ... \\ ... <x_1,y_1|e^{-\varepsilon\widehat{H}}|x_0,y_0> g(x_{n+1}...g_0) \quad (0.11)$$

Here $x_{n+1} = x(\tau = T), y_{n+1} = y(\tau = T), x_0 = x(\tau = 0), y_0 = y(\tau = 0)$. To calculate the integral in equation (1.11), it is necessary to convert the expression for the propagator $<x, y | e^{-\tau \hat{H}} | x', y'>$. Because quantum mechanics is known [5] that

$$<x, y | e^{-\tau \hat{H}} | x', y'> = N(\varepsilon) e^{\varepsilon \hat{L}} \qquad (0.12)$$

In this equation $N(\varepsilon)$ the normalization constant, and $\hat{L}$ is an analogue of the Lagrange operator of quantum mechanics. Expressions for these quantities are given below:

$$N(\varepsilon) = \frac{e^{y(1/2-\alpha)}}{2\pi\varepsilon\xi`\sqrt{1-\rho^2}} \qquad (0.13)$$

$$\hat{L} = -\frac{e^{-y}}{2(1-\rho^2)}(\frac{\delta x}{\varepsilon} + r` - \frac{1}{2}e^y)^2 +$$
$$+\frac{\rho e^{y(1/2-\alpha)}}{\xi`(1-\rho^2)}(\frac{\delta x}{\varepsilon} + r` - \frac{1}{2}e^y)(\frac{\delta y}{\varepsilon} + \lambda` e^{-y} + \mu - \frac{1}{2}\xi` e^{2y(\alpha-1)}) - \qquad (0.14)$$
$$-\frac{e^{2y(1-\alpha)}}{2\xi`^2(1-\rho^2)}(\frac{\delta y}{\varepsilon} + \lambda` e^{-y} + \mu + \frac{1}{2}\xi` e^{2y(\alpha-1)})^2$$

Now we give the obtained Lagrangian $\hat{L}$ to the form $\hat{L} = \alpha_1 A^2(x, y) + \alpha_2 B^2(y)$ in order to break all the integrals on the product of two components. For this we consider the first two terms of equation (1.14):

$$\widehat{L(x, y)} = -\frac{e^{-y}}{2(1-\rho^2)}(\frac{\delta x}{\varepsilon} + r` - \frac{1}{2}e^y)^2 +$$
$$+\frac{\rho e^{y(1/2-\alpha)}}{\xi`(1-\rho^2)}(\frac{\delta x}{\varepsilon} + r` - \frac{1}{2}e^y)^2(\frac{\delta y}{\varepsilon} + \lambda` e^{-y} + \mu - \frac{1}{2}\xi` e^{2y(\alpha-1)}) \qquad (0.15)$$

Passed out of the bracketsthe value $-e^{-y}/2(1-\rho^2)$ we obtain

$$\widehat{L(x, y)} = -\frac{e^{-y}}{2(1-\rho^2)}[(\frac{\delta x}{\varepsilon} + r` - \frac{1}{2}e^y)^2 -$$
$$-\frac{2\rho e^{y(3/2-\alpha)}}{\xi`}(\frac{\delta x}{\varepsilon} + r` - \frac{1}{2}e^y)^2(\frac{\delta y}{\varepsilon} + \lambda` e^{-y} + \mu - \frac{1}{2}\xi` e^{2y(\alpha-1)})] \qquad (0.16)$$

Now we will append the right side of the equation (1.16) to the complete square of the difference:

$$\widehat{L(x,y)} = -\frac{e^{-y}}{2(1-\rho^2)}[(r`-\frac{1}{2}e^y)-\frac{2\rho e^{y(3/2-\alpha)}}{\xi`}(\frac{\delta y}{\varepsilon}+\lambda`e^{-y}+\mu-\frac{1}{2}\xi`e^{2y(\alpha-1)})]^2 -$$
$$-\frac{e^{-y}}{2(1-\rho^2)}\frac{\rho^2 e^{y(3-2\alpha)}}{\xi`^2}(\frac{\delta y}{\varepsilon}+\lambda`e^{-y}+\mu-\frac{1}{2}\xi`e^{2y(\alpha-1)})^2 \quad (0.17)$$

Substituting this expression into equation (1.14), combining similar terms and obtains the required form of the Lagrangian:

$$\hat{L} = -\frac{e^{-y}}{2(1-\rho^2)}[(\frac{\delta x}{\varepsilon}+r`-\frac{1}{2}e^y)-\frac{2\rho e^{y(3/2-\alpha)}}{\xi`}(\frac{\delta y}{\varepsilon}+\lambda`e^{-y}+\mu+\frac{1}{2}\xi`e^{2y(\alpha-1)})]^2 -$$
$$-\frac{e^{2y(1-\alpha)}}{2\xi`^2}(\frac{\delta y}{\varepsilon}+\lambda`e^{-y}+\mu-\frac{1}{2}\xi`e^{2y(\alpha-1)})^2 = \widehat{L(x,y)}+\widehat{L(y)} \quad (0.18)$$

Thus we have an expression for the Lagrangian in the right for further calculations form. The solution of the integral eguation (1.11) for various types of function $g(x,...x`)$ is presented below.

## 4. Path-dependent options of the European type

Path-dependent options, these are options, for which payoff function depends on the entire history of price changes over the term of the option contract. In this work, this means that in contrast to local payoff functions, it is impossible to undertake an analytical integration over $dx_n...dx_1$, so the integration will be carried out numerically. In this case, the general equation for the path-dependent option price takes the form:

$$f(x,y,\tau) = e^{-r`\tau}\int_{-\infty}^{\infty}\prod_{i=0}^{n}dx_i dy_i <x_{i+1},y_{i+1}|e^{-\varepsilon\widehat{H}}|x_i,y_i> g(x_{n+1}...g_0) \quad (0.19)$$

For simplicity of computation we fix some of the parameters of our model: $\lambda = 0, \alpha = 1$. Then the expression for the propagator of the previous equation is rewritten as

$$\left\langle x_{i+1}, y_{i+1} \left| e^{-\varepsilon \widehat{H}} \right| x_i, y_i \right\rangle = \frac{e^{-y_{i+1}/2}}{2\pi\varepsilon\xi\sqrt{1-\rho^2}} \times$$

$$\times \exp\left[ -\frac{e^{-y_{i+1}}}{2(1-\rho^2)} \left[ \left( x_i - x_{i+1} - \varepsilon(r` - \frac{1}{2}e^y) \right) - \frac{2\rho e^{y_{i+1}/2}}{\xi`} \left( y_i - y_{i+1} - \varepsilon(\mu - \frac{1}{2}\xi) \right) \right]^2 - \right.$$

$$\left. - \frac{1}{2\xi^2} \left( y_i - y_{i+1} - \varepsilon(\mu - \frac{1}{2}\xi) \right)^2 \right]$$

Also, as for vanilla European options $N(\varepsilon)$ can be represented as the product of two components:

$$N(\varepsilon) = \frac{e^{y/2}}{2\pi\varepsilon\xi`\sqrt{1-\rho^2}} = \frac{e^{-y/2}}{\sqrt{2\pi\varepsilon\xi`(1-\rho^2)}} \cdot \frac{1}{\xi`\sqrt{2\pi\varepsilon}}$$

Now write equation (1.19) in the form

$$f(x, y, \tau) = e^{-r\tau} \int_{-\infty}^{\infty} DY \rho(Y) F(Y) \qquad (0.20)$$

where

$$DY = \prod_{i=0}^{n} dy_i$$

$$\rho(Y) = \prod_{i=0}^{n} \frac{1}{\xi`\sqrt{2\pi\varepsilon}} \exp\left( -\frac{1}{2\xi^2\varepsilon} \left( y_i - y_{i+1} - \varepsilon(\mu - \frac{1}{2}\xi) \right)^2 \right) \qquad (0.21)$$

$$F(Y) = \int_{-\infty}^{\infty} \prod_{i=0}^{n} \frac{dx_i e^{-y_{i+1}/2}}{\sqrt{2\pi\varepsilon\xi`(1-\rho^2)}} \exp\left( -\frac{e^{-y_{i+1}}}{2(1-\rho^2)\varepsilon} \left[ \left( x_i - x_{i+1} - \varepsilon(r` - \frac{1}{2}e^{y_{i+1}}) \right) - \right. \right.$$

$$\left. \left. - \frac{2\rho e^{y_{i+1}/2}}{\xi`} \left( y_i - y_{i+1} - \varepsilon(\mu - \frac{1}{2}\xi) \right) \right]^2 \cdot g(x_{n+1}, \ldots, x_0) \right) \qquad (0.22)$$

It is seen that $\rho(Y)$ is a function of the density of the joint normal distribution of $y$. Consequently, the integral on the right side of equation (1.20) can be calculated numerically by the Monte Carlo. The easiest way to do it consistently ($n+1$ times) to generate random numbers with distribution (1.21). This method has drawbacks, as each subsequent value $y_{i+1}$ depends on the previous value $y_i$ of this quantity. We present a multidimensional integral as a product of $n+1$ integrals. This will speed up the computation. The idea of partitioning of this integral to the product of several integrals described in [7]. Following this idea we first change the variables in the equation (1.21):

$$y_i = \sqrt{\varepsilon}\xi\eta_i - (\mu - \frac{1}{2}\xi)\varepsilon i, \quad dy_i = \sqrt{\varepsilon}\xi\eta_i.$$

After the change of variables equation (1.20) can be rewritten as:

$$\upsilon(x,\eta_{n+1},\tau) = e^{-r\tau}\int_{-\infty}^{\infty}\prod_{i=0}^{n}\frac{d\eta_i}{\sqrt{2\pi}}\exp\left(-\frac{1}{2}(\eta_i - \eta_{i+1})^2\right)\cdot F(\eta_{n+1},\ldots,\eta_0,x) =$$
$$= e^{-r\tau}\int_{-\infty}^{\infty}\frac{1}{(2\pi)^{\frac{n+1}{2}}}\prod_{i=0}^{n}d\eta_i\exp\left(-\frac{1}{2}(\eta_i - \eta_{i+1})^2\right)\cdot F(\eta_{n+1},\ldots,\eta_0,x) \quad (0.23)$$

Consider the sum in the exponent:

$$\sum_{i=0}^{n}(\eta_i - \eta_{i+1})^2 = \eta_{n+1}^2 - 2\eta_{n+1}\eta_n + \eta_n^2 + \ldots + \eta_1^2 - 2\eta_1\eta_0 + \eta_0^2 =$$
$$= \eta^t M\eta + [\eta_{n+1}^2 - 2\eta_{n+1}\eta_n - 2\eta_1\eta_0 + \eta_0^2] \quad (0.24)$$

Here $\eta$ is n-dimensional vector, $\eta^t$ is the transposed vector, and $M$ is diagonal symmetric matrix of dimension $n \times n$:

$$\eta = \begin{pmatrix} \eta_n \\ \eta_{n-1} \\ \vdots \\ \eta_2 \\ \eta_1 \end{pmatrix}, \quad M = \begin{pmatrix} -2 & -1 & 0 & 0 & 0 \\ -1 & -2 & -1 & 0 & 0 \\ 0 & -1 & \ddots & -1 & 0 \\ 0 & 0 & -1 & -2 & -1 \\ 0 & 0 & 0 & -1 & -2 \end{pmatrix}$$

Changing variables $\eta_i = \sum_{j=1}^{n}O_{ij}\omega_j$, where $O$ is the matrix, which diagonalizes the matrix $M$. In this case we took into account that the Jacobian of such change of variables is equal to 1 by virtue of the orthogonality of the matrix $O$. Equation (1.24), given that the $m_i$ are eigenvalues of the matrix $M$ can be written as:

$$\eta^t M\eta = \omega^t O^t MO\omega = \omega^t M_d\omega = \sum_{i=1}^{n}m_i\omega_i^2 \quad (0.25)$$

Substitute equation (1.25) in equation (1.24), and provide allocation of the full square of the differencea. As a result, we obtain the desired representation for the equation (1.24):

$$\sum_{i=0}^{n}(\eta_i - \eta_{i+1})^2 = \sum_{i=0}^{n}m_i[\omega_i - \frac{(\eta_0 O_{1i} - \eta_{n+1}O_{ni})}{m_i}]^2 + \eta_{n+1}^2 + \eta_0^2 -$$
$$- \sum_{i=0}^{n}\frac{(\eta_0 O_{1i} - \eta_{n+1}O_{ni})^2}{m_i}$$

Using this expression for the sum in the exponent of equation (1.23), we rewrite this equation in the form:

$$\upsilon(x, y_{n+1}, \tau) = e^{-r\tau} \int_{-\infty}^{\infty} dy_0 \cdot g(y_{n+1}, y_0) \cdot$$

$$\cdot \int_{-\infty}^{\infty} \frac{1}{(2\pi)^{\frac{n+1}{2}}} \prod_{i=1}^{n} d\omega_i \left\{ \exp\left(-\frac{1}{2} \sum_{i=1}^{n} m_i \left(\omega_i - \frac{\vartheta_i}{m_i}\right)^2\right) \right\} \cdot F(y_{n+1}, \omega_{n-1}, \ldots, \omega_1, y_0, x) \quad (0.26)$$

$$\vartheta_i = (\eta_0 O_{1i} - \eta_{n+1} O_{ni}) = \frac{y_0 O_{1i} - [y_{n+1} + (\mu - \frac{1}{2}\xi)(n+1)\varepsilon] O_{ni}}{\sqrt{\varepsilon \xi}}$$

In this equation, we have collected in the function $g(y_{n+1}, y_0)$ of everything that depends on $y_0$ and $y_{n+1}$. The function $g(y_{n+1}, y_0)$ in this case is given by:

$$g(y_{n+1}, y_0) = \frac{1}{\sqrt{2\pi\varepsilon\xi}} \exp\left( \frac{-[(y_{n+1} + (\mu - \frac{1}{2}\xi)(n+1)\varepsilon)^2 + y_0^2 - \varepsilon\xi^2 \sum_{i=0}^{n} \frac{\vartheta_i^2}{m_i}]}{2\varepsilon\xi^2} \right)$$

Let us make one more change of variables: $\omega_i = \frac{\vartheta_i}{m_i} + \frac{\zeta_i}{\sqrt{m_i}}$. After this change of variables, equation (1.26) reads as:

$$\upsilon(x, y_{n+1}, \tau) = e^{-r\tau} \int_{-\infty}^{\infty} dy_0 \cdot g(y_{n+1}, y_0) \cdot \int_{-\infty}^{\infty} \prod_{i=1}^{n} d\zeta_i \frac{1}{\sqrt{2\pi}} \exp\left(-\frac{\zeta_i^2}{2}\right) \tilde{F}(y_{n+1}, \zeta_n, \ldots, \zeta_1, y_0)$$

.

In this equation, the following equality takes place:
$$\tilde{F}(y_{n+1}, \zeta_n, \ldots, \zeta_1, y_0) = F(y_{n+1}, y_n, \ldots, y_1, y_0),$$

if one makes the substitution

$$y_i = \sqrt{\varepsilon\xi} \sum_{j=0}^{n} O_{ij} \left( \frac{\zeta_j}{\sqrt{m_j}} + \frac{\vartheta_j}{m_j} \right) - (\mu - \frac{1}{2}\xi)\varepsilon i \quad (0.27)$$

This substitution is obtained if one made to restore all the change of variables and to reverse these changes of variables. It should be note that when we made the change of variables $\omega_i = \frac{\vartheta_i}{m_i} + \frac{\zeta_i}{\sqrt{m_i}}$, the Jacobian of such change of variables is equal to $1/\sqrt{m_i}$. In total the Jacobian of all such substitutions will give the product for all $\prod_{i=1}^{n} 1/\sqrt{m_i} = 1/\sqrt{\det M}$. For convenience, we include the contribution from the Jacobian of these substitutions in the function $g(y_{n+1}, y_0)$:

$$g(y_{n+1}, y_0) = \frac{1}{\sqrt{2\pi\varepsilon \det M}\, \xi} \exp\left( \frac{-[(y_{n+1} + (\mu - \frac{1}{2}\xi)(n+1)\varepsilon)^2 + y_0^2 - \varepsilon\xi^2 \sum_{i=0}^{n} \frac{\vartheta^2}{m_i}]}{2\varepsilon\xi^2} \right)$$

Now are made all the necessary transformations, which allowed us to break out of the integral from equation (1.20) on the outer integral over $dy_0$ and a product of $n$ integrals over $dy_i$, $i \in [1,n]$. Further we shall calculate the outer integral by means of one of the quadrature formulas, for example Simpson's formula. To do this, we represent equation (1.20) in the form:

$$f(x, y_{n+1}, \tau) = e^{-r\tau} \int_{-\infty}^{\infty} dy_0 g(y_0) \chi(y_0)$$
$$\chi(y_0) = \int_{-\infty}^{\infty} \prod_{i=1}^{n} d\varsigma_i \rho(\varsigma_i) \widetilde{F}(y_{n+1}, \varsigma_n ... \varsigma_1, y_0), \quad (0.28)$$

where $\rho(\varsigma_i)$ is the density function of normal distribution with the parameters $N(0,1)$. To calculate function $\chi(y_0)$ we use the Monte-Carlo method. The implementation of this method is described below in section 5.

## 5. Monte-Carlo approach to path-dependent options

To calculate the quantity $\chi(y_0)$ from the equation (1.28) with the Monte Carlo method, we proceed as follows. Since the probability density function of $y_i$ in this case does not depend on previous values of the quantity $y_{i+1}$, one needs to generate $n$ random numbers with normal distribution $N(0,1)$. Then we have to calculate all the values of $y_i$ accoding to the formula (1.27), and thus generate the entire trajectory $(y_{i+1}, y_n, ..., y_1, y_i)$. Then one has to substitute this expression for the trajectory into the function $F(y_{n+1}, y_n ... y_1, y_0)$. Now, in order to calculate the price of path-dependent option, it is necessary to calculate the value of function $F(y_{n+1}, y_n ... y_1, y_0)$:

$$F(Y) = \int_{-\infty}^{\infty} \prod_{i=0}^{n} \frac{dx_i \cdot e^{-y_{i+1}/2}}{\sqrt{2\pi\varepsilon\xi`(1-\rho^2)}} \times \exp\left\{ \frac{-e^{y_{i+1}}}{2(1-\rho^2)\varepsilon}[x_i - (x_{i+1} + \varepsilon(r` - \frac{1}{2}e^{y_{i+1}})) - \frac{2\rho e^{y_{i+1}/2}}{\xi`}(y_i - (y_{i+1} + \varepsilon(\mu - \frac{1}{2}\xi`)))]^2 \right\} g(x_{n+1}...x_0)$$

Below we give the expression for the function $F$ to a form suitable for the application of the Monte-Carlo metod:

$$F(Y) = \int_{-\infty}^{\infty} \prod_{i=0}^{n} dx_i \rho(x_{i+1}, x_i) g(x_{n+1}...x_0)$$

$$\rho(x_{i+1}, x_i) = \frac{e^{-y_{i+1}/2}}{\sqrt{2\pi\varepsilon\xi`(1-\rho^2)}} \times \exp\left\{ \frac{-e^{y_{i+1}}}{2(1-\rho^2)\varepsilon}[x_i - (x_{i+1} + \varepsilon(r` - \frac{1}{2}e^{y_{i+1}})) - \right.$$

$$\left. -\frac{2\rho e^{y_{i+1}/2}}{\xi`}(y_i - (y_{i+1} + \varepsilon(\mu - \frac{1}{2}\xi`)))]^2 \right\}$$

Also we write the formula for generating random numbers with distribution, necessary for the Monte Carlo calculations:

$$x_i = x_{i+1} + \varepsilon(r` - \frac{1}{2}e^{y_{i+1}})) + \frac{2\rho e^{y_{i+1}/2}}{\xi`}(y_i - (y_{i+1} + \varepsilon(\mu - \frac{1}{2}\xi`))) + \\ + e^{y_{i+1}/2}\sqrt{\varepsilon\xi`(1-\rho^2)}N(0,1) \quad (0.29)$$

Now, as above, it necessary to generate a trajectory $(x_{i+1}, x_n, ..., x_1, x_i)$, and then substitute it to the payoff function $g(x_{n+1}...x_0)$ for path-dependent options. We write the final algorithm for computing the function $\chi(y_0)$ of equation (1.28):

1. For each value of $y_0$ generate $N$ trajectories $Y$ accoding to (1.29);
2. Substitute each trajectory $Y$ into $F(Y)$ and generate trajectories $X$;
3. Calculate function $g(X)$ for all trajectories;
4. Calculate mean value of $g(X)$ over all trajectories.

## 6. Results of the numerical calculations

In this section we present some computational results obtained using the methods described above. Since the value of implied volatility plays an important role in options pricing, it will be necessary to examine how the parameters of stochastic volatility models affect the graph of Implyed volatility. Figures 1,2,3,4 show graphs of implied volatility as a function of strike price for different values of $\xi, \rho, k$. In calculating Imlied Volatility the following parameters of Call option were used: $S = 0$, $T = 1$, $r = 0.04$, $V_0 = 0.3$.

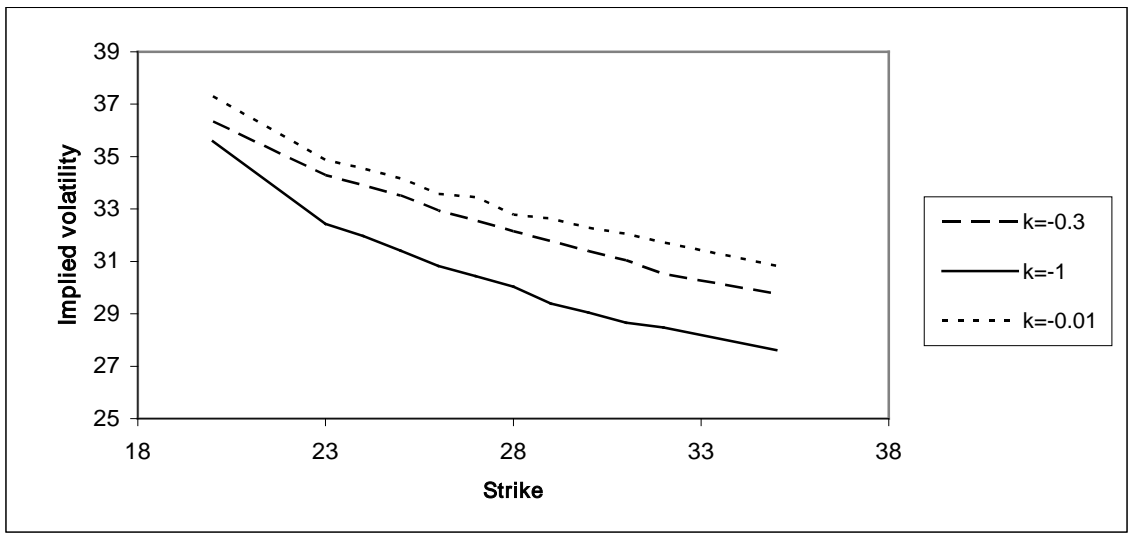

**Figure 1. Graph of implyed volatility for various values of k.**

Note that the "rate of return to the mean" $k$ effect on the height of the curve Implyed Volatility, but did not change either the slope or curvature

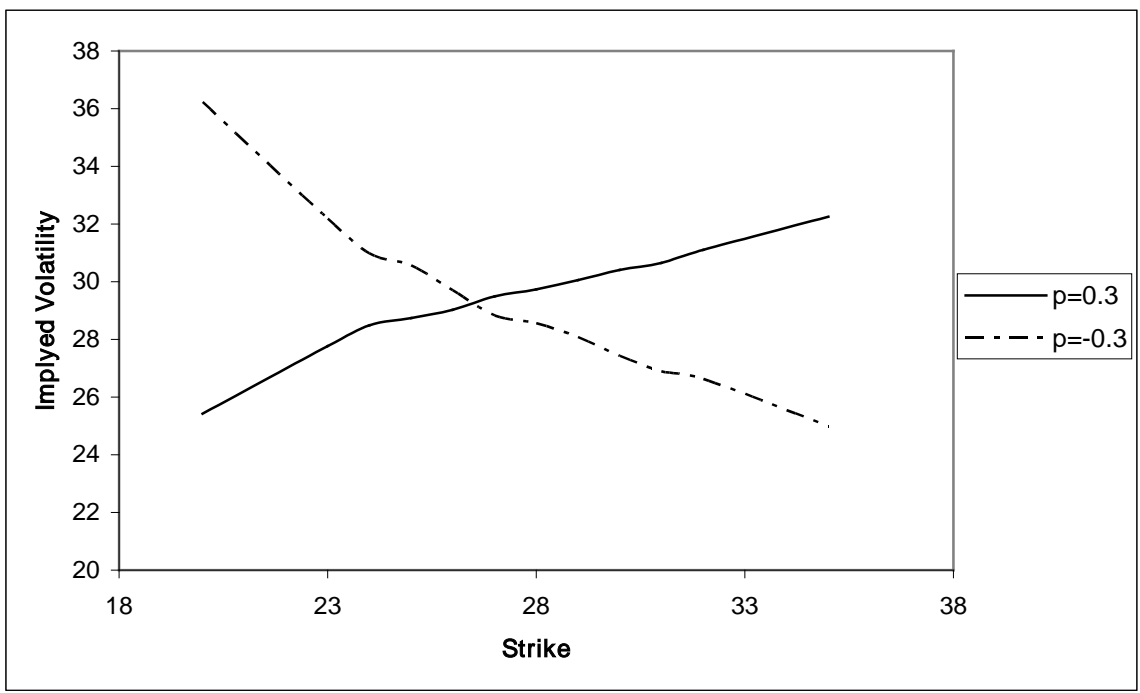

**Figure 2. Graph of implied volatility for various values of the correlation parameter.**

Figure 2 shows that the correlation parameter has a very strong influence on the graphic of Imlied Volatility.

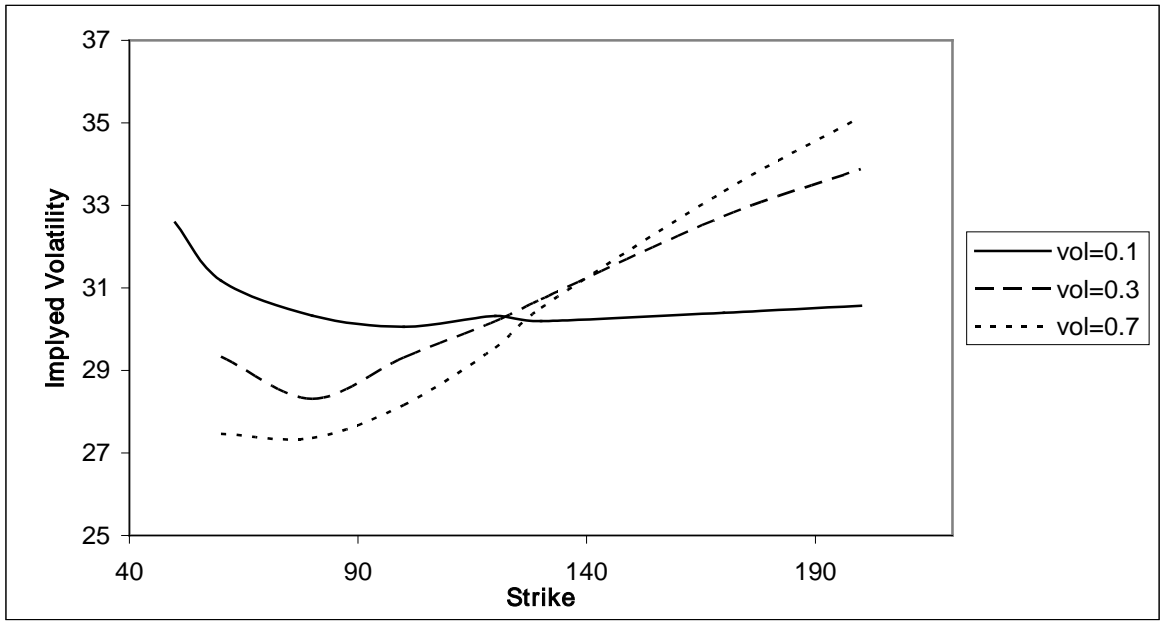

**Figure 3. Graph of implied volatility for various values of the volatility of volatility parameter.**

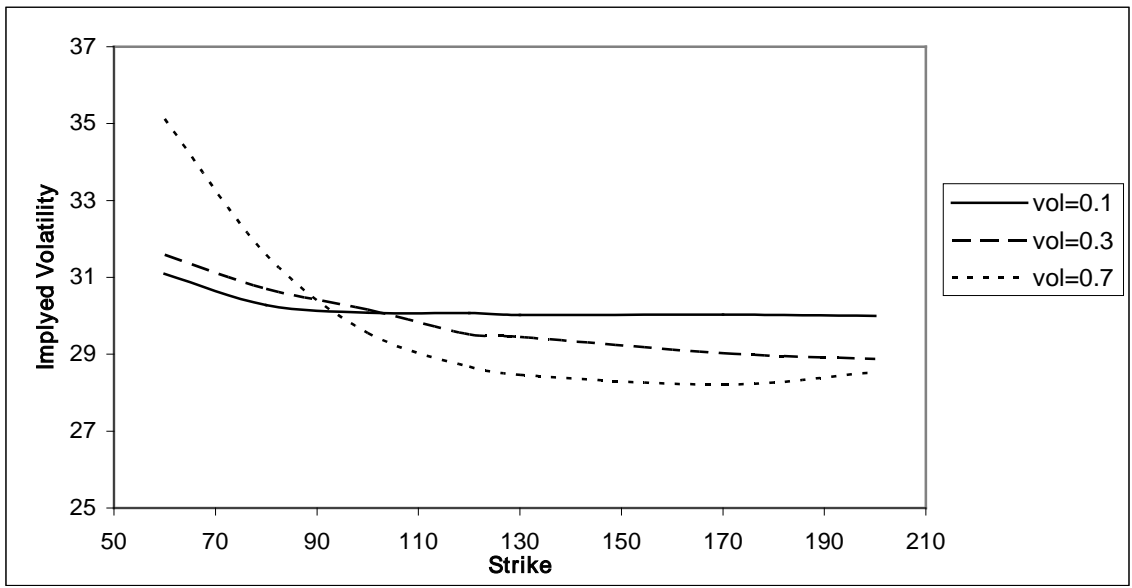

**Figure 4. Graph of implied volatility for various values of the volatility of volatility parameter.**

Figures 3, 4 present graphs of implyed volatility for various values of the volatility of volatility parameter. Implyed volatility values were calculated for correlation coefficients $\rho = 0.3$ and $\rho = -0.3$ respectively. From these figures we can conclude that the volatility of volatility parameter $\xi$ changes the angle of inclination of the implyed volatility graph.

Figures 1,2,3,4 show that changing the parameters of the stochastic volatility model one can significantly change the implied volatility function. From this it follows that, given the right fitting in the model parameters, the method described here will provide close to to real values of Imlied Vilatility and therefore the option prices close to real values. Figure 5 shows a graph of Imlied Volatility, calculated on the basis of real data, calculared using the method of stochastic volatility and calculated at the constant volatility.

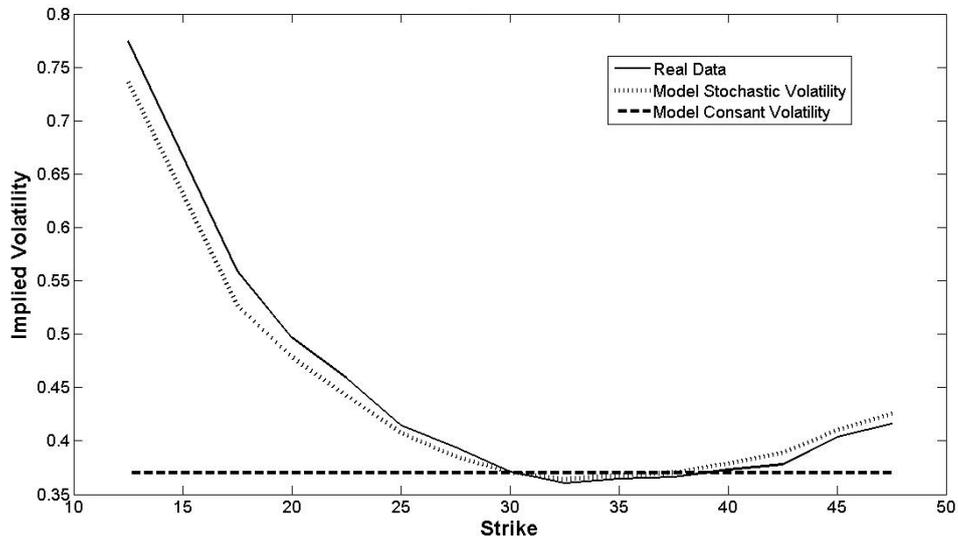

**Figure 5. Graphs of Implied Volatility: comparison of the real and model data.**

The figure 5 shows that the model described in this paper allows a fairly accurate description of the implyed volatility function. To implement these methods the software in Java has been written. The outer integral in equation (1.28) was calculated using the trapezoid method with the number of partition points is equal to 100. The inner integral is calculated by the Monte Carlo with the number of trajectories of 1000. Function $F(Y)$ was calculated by the Monte-Carlo by generating 10,000 paths.

Summary

In this paper we developed analytical and numerical methods for assessing path-dependent options of European type. As a model describing dynamics of the underlying asset volatility was chosen as a model of stochastic volatility [5,8]. This allowed a fairly accurate description of the implied volatility function. This is an important result, since a correct description of the implied volatility function can take into account the effect of "volatility smile" in options pricing. This allows considerably improve the accuracy of the calculation of the options prices and bring the theoretical prices to the real prices of the options trading. Accounting for the effect of "volatility smile" is one of the most important issues in the options evaluating. The results of this study showed that the use of a stochastic volatility model gives good results in addressing the volatility smile. This means that developed method can be applied in a practical option trading. Although the results of [3,5,7,8,10,11,12,13], which used the model of stochastic volatility and the method of path integral, are well known, we were generalized ideas and results of these articles on the case of the European type options, in which the payoff function depends on the values of the underlying asset for the entire duration of the option.

Our method is universal in the sense that it can be applied without changes to any payoff functions of the European type options. Although the rate of calculations in this method is slow in comparison with the vrate of calculations from the exact formulas, but, as mentioned above, the main advantage of yhis method is its universality with respect to the payoff functions. In addition, options with complicated payoff functions are traded infrequently, and, consequently, the rate of calculations for these options is not critical. Thus we can conclude that our method can be practically relevant for evaluating European options with complicated payoff functions.